\documentclass[journal]{IEEEtran}

\usepackage{cite}
\usepackage{amsmath,amssymb,amsfonts}
\usepackage{graphicx}
\usepackage{afterpage}
\usepackage{booktabs}
\usepackage{array}
\newcolumntype{L}[1]{>{\raggedright\arraybackslash}p{#1}}
\usepackage{url}
\usepackage{xurl} 
\usepackage{xcolor}
\usepackage[hidelinks,hypertexnames=false]{hyperref}

\usepackage{algorithm}
\usepackage{listings}
\usepackage{upquote}
\definecolor{clrKey}{HTML}{0B5394}    
\definecolor{clrCmt}{HTML}{4C7A5D}    
\definecolor{clrData}{HTML}{0F766E}   
\definecolor{clrCtrl}{HTML}{B45309}   
\definecolor{clrCfg}{HTML}{7C3AED}    
\definecolor{clrCoef}{HTML}{BE123C}   
\definecolor{clrPwr}{HTML}{1D4ED8}    
\definecolor{clrOp}{HTML}{6B7280}     
\definecolor{clrBg}{HTML}{FAFBFD}     
\definecolor{clrRule}{HTML}{9AA7B8}   
\newcommand{\lstcmt}{\itshape\color{clrCmt}}

\lstdefinestyle{pseudoverilog}{%
  language=Verilog,
  basicstyle=\ttfamily\scriptsize\color{black},
  keywordstyle=\bfseries\color{clrKey},
  commentstyle=\lstcmt,
  morecomment=[l]{//},
  alsoletter={_0123456789},
  emph={[1]data,data_hd,data_bin,HD},
  emphstyle={[1]\color{clrData}},
  emph={[2]vld,rdy,mult},
  emphstyle={[2]\color{clrCtrl}},
  emph={[3]mask,a_mode,sel0,sel1,sel2},
  emphstyle={[3]\color{clrCfg}},
  emph={[4]coef0,coef1,coef2,bias,leaf_coef},
  emphstyle={[4]\color{clrCoef}},
  emph={[5]prxy0,prxy1,prxy2,pwr},
  emphstyle={[5]\bfseries\color{clrPwr}},
  literate={\&}{{\color{clrOp}\&}}1
           {|}{{\color{clrOp}|}}1
           {~}{{\color{clrOp}\textasciitilde}}1
           {*}{{\color{clrOp}*}}1
           {+}{{\color{clrOp}+}}1
           {=}{{\color{clrOp}=}}1
           {<}{{\color{clrOp}<}}1
           {>}{{\color{clrOp}>}}1,
  backgroundcolor=\color{clrBg},
  columns=fullflexible,
  keepspaces=true,
  showstringspaces=false,
  numbers=left,
  numberstyle=\tiny\color{clrOp},
  numbersep=5pt,
  xleftmargin=1.6em,
  framexleftmargin=1.6em,
  framexrightmargin=0.2em,
  aboveskip=2pt,
  belowskip=0pt,
  breaklines=false,
  frame=none,
  escapeinside={(*@}{@*)},
}

\begin{document}

\title{X-OPM: Explainable Automatic Digital On-Chip Power Modeling for Enhanced Robustness}

\author{Jingbo~Jiang,
        Xizi~Chen,
        Jian~Peng,
        and~Wei~Zhang%
\thanks{J.~Jiang, J.~Peng, and W.~Zhang are with the Department of Electronic and Computer Engineering, The Hong Kong University of Science and Technology, Hong Kong (e-mail: jjiangan@connect.ust.hk; jpengai@connect.ust.hk; wei.zhang@ust.hk).}%
\thanks{X.~Chen is with Huazhong Agricultural University, Wuhan, China (e-mail: xchenbn@mail.hzau.edu.cn).}}

\markboth{Preprint}%
{Jiang \MakeLowercase{\textit{et al.}}: X-OPM: Explainable Automatic Modeling for On-Chip Power Meters}

\maketitle

\begin{abstract}
Proactive power management systems reduce processor dynamic power through runtime power prediction and power-aware scheduling. Accurate, stable and low-overhead digital on-chip power meters (OPMs) are crucial for improving the prediction quality. Recent studies have explored various modeling methods, including using linear models, decision trees, and multi-layer perceptrons (MLPs) to construct OPMs. However, most current approaches train models end-to-end without analyzing the physical interpretability of features, affecting their ability to generalize to unseen workloads. Grounded in the design principles of synchronous digital VLSI circuits, X-OPM introduces a robust feature engineering framework that uses tree-based models to capture feature interactions and linear models for prediction. It also incorporates a human-in-the-loop workflow to balance model accuracy against modeling effort. Evaluated on a commercial C906 vector processor, X-OPM consistently achieves $R^2 > 0.93$ across all workloads with sampling window size set below $8$ cycles. In contrast, state-of-the-art methods including APOLLO, COBIT, and standard MLPs fail to generalize across all test cases. Layout with commercial EDA tools shows that X-OPM incurs an area overhead below $0.1\%$, which is on par with lightweight tree-based and linear models, and significantly smaller than MLP-based models.
\end{abstract}


\IEEEpeerreviewmaketitle

\section{Introduction}
\begin{figure}[!t]
\centering
\includegraphics[width=\columnwidth]{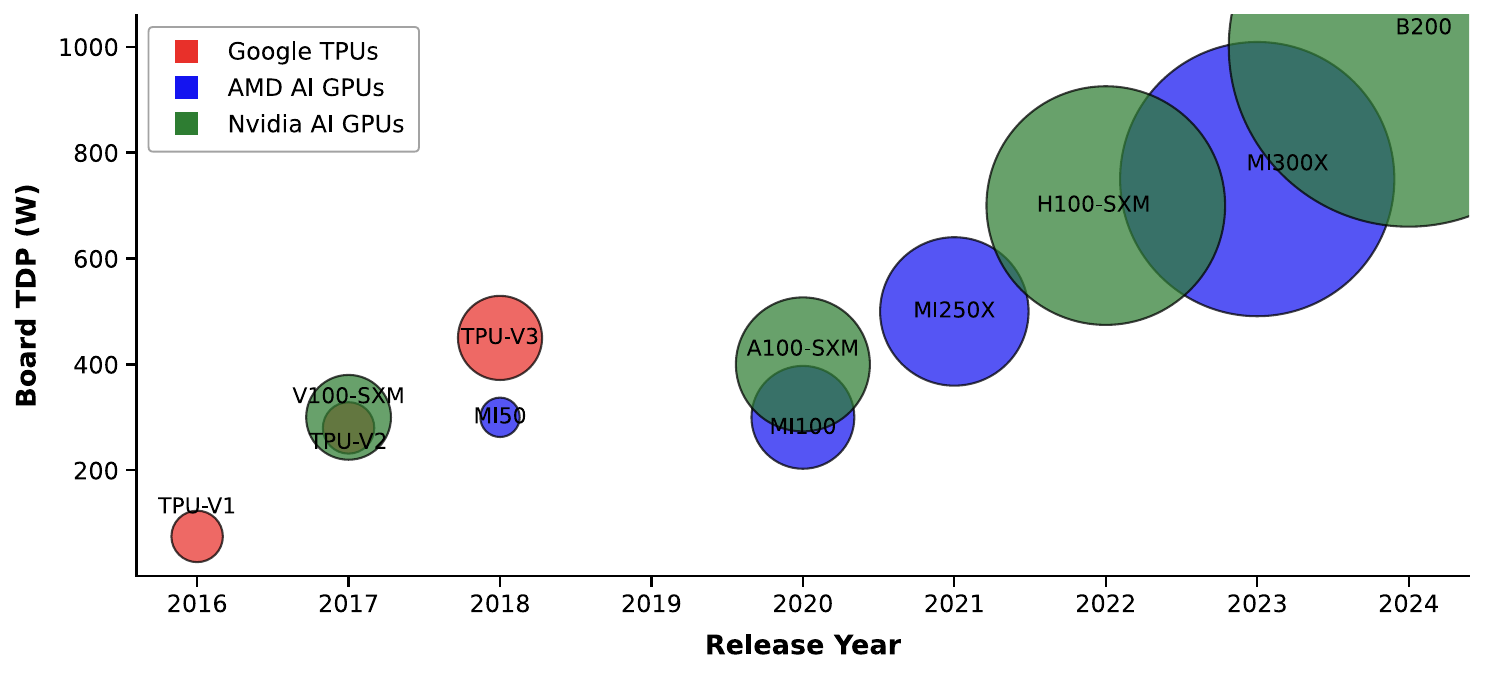}
\caption{Board TDP (Thermal Design Power) versus release year of recent cloud AI processors.}
\label{fig:tdp_trend}
\end{figure}

\IEEEPARstart{T}{he} computational throughput and power consumption of processors, especially GPUs (Graphics Processing Units) and NPUs (Neural Processing Units) running AI benchmarks, have increased drastically in recent years, as illustrated in Fig.~\ref{fig:tdp_trend}. Excessive power consumption incurs a multitude of power and thermal related issues, including sudden cut-offs due to supply voltage fluctuations (di/dt), chip malfunction due to overcurrent, and runtime frequency drops due to overheating~\cite{grochowski2002didt,powell2002damping,joseph2003voltage,leng2014gpuvolt,leng2015gpunoise,thomas2016emergpu,paul2013cooperative}.

To address these challenges, proactive power management systems~\cite{ibmz15proactive,qualcomm2021hexagon,lefurgy2017power9,ibmz14power,proactive65nm2024,ondie2025runtime,isscc2026proactive} have been developed to improve processor runtime stability and maximize performance. A proactive power management system employs a digital on-chip power meter (OPM) to predict dynamic power fluctuations of the processor at runtime. The power predictions can typically be made several tens of cycles in advance with an accuracy (measured in $R^2$) of over $0.9$. When the processor anticipates an imminent power event, it proactively mitigates the impact by performing runtime scheduling, such as instruction throttling or adaptive clock gating.

Different methods have been proposed to build OPMs, spanning from using the linear model with feature selection APOLLO~\cite{xie2021apollo} and DEEP~\cite{zhai2022deep}, to extracting high-dimensional features using non-linear models such as trees~\cite{li2025cobit} or multi-layer perceptrons (MLPs)~\cite{proactive65nm2024,isscc2026proactive}. Although current methods focus on trying different models, limited attention has been drawn to improving the modeling robustness. The model robustness is defined as the ratio of the $R^2$ score an OPM can achieve on unseen workloads to that on the training dataset. 

Clarifying the model robustness before tape-out is crucial for preventing OPM from malfunction after the chip comes back. OPM typically implements the (\textit{features}), also called the \textit{power proxies}, as combinational circuits hardened on-chip that cannot be modified after tape-out. OPM relies on the accurate post-layout power simulation or emulation to collect the \textit{labels} for training the model in the pre-silicon stage. The OPM is typically fitted within a short timeframe of several months, spanning from RTL (Register Transfer Level) freeze to tape-out. Slow post-layout simulation constrains the training data to a small set of testbenches instead of all possible workloads. High-dimensional models such as trees and MLPs are prone to overfitting when the training dataset is small, which can lead to a decrease in model robustness.

To improve the model robustness, we propose X-OPM, an automatic white-box power modeling framework that integrates the prior knowledge of RTL design into the model fitting pipeline. X-OPM first categorizes all RTL signals into three categories -- the control, configuration, and the data signals, then applies unique signal post-processing and feature interaction algorithms to each category. The signals are transformed to be positively and linearly correlated to the power before being fitted with a regularized linear model. Through several iterations of adding and removing signals, the relationship between each proxy and power becomes fully explainable. This transparency ensures the model can be trusted to generalize to unseen workloads.

We evaluate X-OPM on the XuanTie C906~\cite{c906datasheet}, a RISC-V core with a 128-bit vector unit, laid out in TSMC 28\,nm and labelled by sign-off power simulation. X-OPM holds $R^2 > 0.93$ on every test workload at sampling windows below $8$ cycles. As a comparison, OPMs built with APOLLO, COBIT, and the MLP~\cite{proactive65nm2024} approach fail to maintain test accuracy on a different test set. The layout result shows that the area overhead of our OPM is under $0.1\%$ of the C906 core, on par with the linear and tree model and far below the MLP model. Our contributions are:
\begin{itemize}
\item Proposed a white-box OPM modeling methodology derived from RTL design principles. This method ensures strict feature explainability and improves the model's robustness to unseen workloads.
\item Proposed a hybrid model backbone matched to this methodology, in which a tree model constructs the feature interactions and a linear model performs the prediction.
\item Proposed a scalable, automated framework for building the OPM based on the proposed white-box modeling methodology.
\end{itemize}

The rest of this paper is organized as follows. Section~\ref{sec:background} reviews on-chip power modeling methods. Section~\ref{sec:methodology} discusses the proposed X-OPM white-box modeling method. Section~\ref{sec:auto} presents an automated model building flow based on the proposed white-box method. Section~\ref{sec:experiments} compares the C906 OPM's modeling results using X-OPM with baselines including APOLLO, COBIT, and~\cite{proactive65nm2024}. The hardware overhead of X-OPM is also reported.

\section{Background and Related Works}
\label{sec:background}
The power model is used for predicting the dynamic power consumption of a chip running different workloads. It is also referred to as the current model since it is usually trained with the power simulation assuming a fixed PVT (Process Voltage Temperature) corner. The power model can be classified into the on-chip and the off-chip models. OPM, implemented inside the proactive power management system on-chip, is used for predicting the runtime average power of the chip within tens of several cycles. The design of the OPM requires trading off the model accuracy with the implementation overhead, which favors the use of light-weight models such as the look-up-table, linear or tree models.

The off-chip power model, on the other hand, is used for off-line architectural design space exploration~\cite{li2009mcpat, zhou2019primal}. Complex models including deep neural networks~\cite{zhang2020grannite} are the best choices for building the off-chip power models. Other models including the PDN (power delivery network) model~\cite{qualcomm2021hexagon} converting the current to the voltage, or models that predict the PVT change~\cite{isscc2026proactive} are used together with power model in the proactive power management system. The methodology proposed in this paper focuses exclusively on the on-chip power modeling and may not be suitable for other types of models.

Recent OPM research focuses more on modeling large-scale, high-power processors with automated flows. PRIMAL~\cite{zhou2019primal} proposes to fit the model with proxies selected from all RTL module inputs. APOLLO~\cite{xie2021apollo} and DEEP~\cite{zhai2022deep} propose different automated feature selection algorithms for the linear model. Lin et al.~\cite{lin2019ensemble} utilizes decision trees to capture the feature interactions of the on-chip signals. MLPs~\cite{proactive65nm2024,isscc2026proactive}, although incurring significant area overhead, have also been proposed to boost the model accuracy of OPM. 

The literature has found that mixing some circuit design knowledge into the feature engineering procedures helps to improve the accuracy of the power model. PROPHET~\cite{prophetplus2024} proposes to build special proxies that calculate the bit toggle rate of the data buses, and demonstrates that training solely with these special proxies can converge the power model training for the tensor core implemented inside the AI processor. COBIT~\cite{li2025cobit} reasons that the binary tree is one of the most suitable model backbone for fitting the power proxies selected from all signals on-chip, because the joint activity of the binary signals can be represented by the branching of the trees. 

Although all literature has reported a high fitting score, normally $R^2>0.9$, on different benchmarks and under different conditions, discussions about the correlation between the trained signals and the power remain scarce. 

A linear flow never investigates whether a selected net is genuinely causal (e.g., a clock enable, a mode select) or merely an unrelated bus that happened to remain active during the benchmark. A tree flow never verifies if a learned conjunction represents a real hardware gating condition. This purely empirical selection is brittle, exposing the OPM precisely where it is most vulnerable: the dataset is small, the state-space coverage is entirely dependent on verification testbenches, and the final model is permanently frozen at tape-out. Section~\ref{sec:methodology} supplies this missing evidence, leveraging the low-power design rules of synchronous circuits to define structural requirements that guarantee a proxy's correlation with power is causal rather than accidental.

In statistical learning, it is well-known that modeling small models on small datasets requires strong feature engineering fitted with simple but robust model backends to observe and avoid the overfitting. In this paper, we aim to push the white-box modeling method further and propose an automated white-box modeling method that encodes a more comprehensive RTL design knowledge into the automated modeling procedure.

\section{X-OPM Methodology}
\label{sec:methodology}
This section derives the X-OPM model from the low-power design rules of synchronous digital VLSI designs (short as RTL designs). We first analyze the source of dynamic power based on a typical synchronous circuit, then propose methods to construct the white-box model for generalized pipelined RTL designs. 

\begin{figure}[!t]
\centering
\includegraphics[width=\columnwidth]{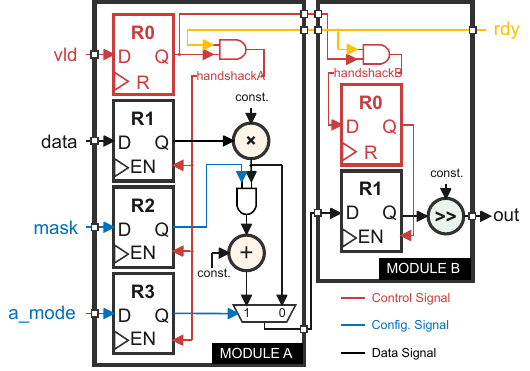}
\caption{A typical single-stage synchronous circuit design for analyzing the source of dynamic power.}
\label{fig:toy_design}
\end{figure}

\subsection{Dynamic Power in Synchronous Circuit}
Fig.~\ref{fig:toy_design} shows a representative synchronous RTL design. It consists of two modules -- Module~A and Module~B, with pipeline registers inserted at its input. Module~A can be configured to perform multiplication or multiply-accumulate (MAC), controlled by the \texttt{a\_mode} signal. Module~B is a fixed constant shifter. The handshake signals \texttt{handshakeA} and \texttt{handshakeB} are used to synchronize the data flow between the two modules. The \texttt{mask} signal is a data-gating signal that controls whether the output of Module~A is passed to Module~B. 

The signals inside an RTL design can be typically classified into three categories: \textbf{control signals}, \textbf{configuration signals} and \textbf{data signals}. Control signals are used to control the start and stop of the module, configuration signals are used to configure the operational mode of the logic, while data signals are typically wide buses that carry numerical data to be processed. In Fig.~\ref{fig:toy_design}, the \texttt{vld}, \texttt{rdy}, \texttt{handshakeA}, and \texttt{handshakeB} signals are control signals since they enable the pipeline registers to pass the data to the downstream. The \texttt{a\_mode} is a configuration signal that determines whether Module~A should operate as a multiplier or a MAC. The \texttt{data} is a data signal. It is worth mentioning that some signals also serve as \textbf{low-power control signals}. RTL designs typically utilize clock gating and data gating techniques to reduce dynamic power. In Fig.~\ref{fig:toy_design}, the handshake signals serve as the clock gating signals that turn on and off the clock of pipeline registers, while the \texttt{mask} signal serves as data-gating signal masking the output of the multiplier so it does not toggle the adder when Module~A is configured to be a MAC. 

Our proposed white-box model is built based on the four categories. Unlike COBIT~\cite{li2025cobit} or APOLLO~\cite{xie2021apollo} that perform feature selection and model regression on all internal nets of a design, we manually select the signals that can be clearly mapped to one of the four categories, discard the rest, and perform post-processing. The processed signals are then fed to the feature selection, feature interaction and model regression pipeline. We typically request RTL designers to provide the list of signals that can be mapped to one of the three categories, or request specific signal naming conventions, such as \texttt{*\_stall} and \texttt{*\_busy}, to be applied across the design to enable automatic signal identification. The signal categories are summarized in Table~\ref{tab:signal_categories}.

\begin{table}[!t]
\centering
\caption{The proposed three-category signals.}
\label{tab:signal_categories}
\renewcommand{\arraystretch}{1.2}
\setlength{\tabcolsep}{3pt}
\footnotesize
\begin{tabular}{@{}L{0.135\columnwidth}L{0.17\columnwidth}L{0.33\columnwidth}L{0.285\columnwidth}@{}}
\toprule
\textbf{Category} & \textbf{Naming Convention} & \textbf{Post-processing}\textsuperscript{*} & \textbf{Power Behavior} \\
\midrule
Control &
\texttt{*\_busy} \newline
\texttt{*\_stall} \newline
\texttt{*\_clk\_en} &
Split signals into single-bit and invert the one that is negatively correlated with power &
Positively correlated with power \\
\addlinespace[3pt]
Config. &
\texttt{*\_mode} \newline
\texttt{*\_sel} \newline
\texttt{*\_cfg} &
Decimal to one-hot &
Non-linear and requires feature interaction \\
\addlinespace[3pt]
Data &
\texttt{*\_data} \newline
\texttt{*\_bus} &
Calculates Hamming distance ($\mathrm{HD}$) between consecutive cycles &
$\mathrm{HD}$ is positively correlated with power  \\
\bottomrule
\multicolumn{4}{@{}l@{}}{\textsuperscript{*}\footnotesize Binning can be applied to all categories.} \\
\end{tabular}
\end{table}

\subsection{Signal Post-processing and Feature Interaction}
\label{sec:postproc}

Different types of signals require distinct post-processing methods. Control signals are transformed to be positively correlated with the power so that the coefficients of the control signals represent the average power increased when the signal is asserted. Multi-bit control signals using one-hot encoding are split into single-bit features. Configuration signals using decimal encoding are first converted to one-hot representations and then decomposed into single-bit features. For instance, a 2-bit \texttt{mode} signal indicating multiplier mode ($\text{decimal } 2$) or MAC mode ($\text{decimal } 3$) is separated into two binary features: $\mathrm{mode}=2$ and $\mathrm{mode}=3$. The coefficient of each feature then denotes the average power consumed in that operating mode. Data signals ($\mathbf{d}$) are transformed to the bitwise Hamming distance ($\mathrm{HD}$) between consecutive clock cycles ($t$ and $t-1$), represented by
\begin{equation}
\label{eq:hd}
\mathrm{HD}[t] = ||\mathbf{d}[t] \oplus \mathbf{d}[t-1]||_1,
\end{equation}
where $\oplus$ denotes the bitwise XOR operation and $\Vert{}\cdot\Vert{}_1$ represents the popcount operation (counting $1$). Because $\mathrm{HD}[t]$ measures the number of bit toggles per cycle, it correlates positively with dynamic power consumption.

\emph{Binning} is a post-processing technique that can be applied to all three categories of signals. Binning collapses a group of signals that generate similar magnitudes of power into a single signal. It decreases the model accuracy but reduces the number of bits required to build the signal. For example, a module that can be configured into $8$ different modes requires an $8$-bit one-hot signal to be modeled precisely. Each one-hot bit is paired with a learnable coefficient $\beta[i]$ representing the average power of that mode. Eight coefficients are required for modeling. Alternatively, these modes can be divided into $8$ bins represented by a $4$-bit signal \texttt{mode\_bin} and one learnable coefficient $\beta$. $mode\_bin==0$ is left unused since the power would be arithmetically computed as $0$. The power of all modes is thus forced into an arithmetic progression, meaning that $mode\_bin==i$ yields a power of $\beta \cdot i$. Binning can also be applied to data signals. For instance, the $\mathrm{HD}$ of a $4096$-bit data signal can be divided into two bins, $\mathrm{HD} < 2048$ and $\mathrm{HD} \geq 2048$, reducing a $13$-bit feature to $2$-bit.

Feature interactions among the three signal categories exist pervasively. For example, modeling the dynamic power of a clock-gated module configured to a specific mode operating on a sequence of data requires the use of \emph{multiplicative interaction}
\begin{equation}
\text{Multiplicative Feature} = \text{Control} \cdot \text{Configuration} \cdot \text{Data}
\end{equation}
Another common type is \emph{selective interaction}. For example, when a configuration signal is asserted, the module operates on data A, otherwise it operates on data B. This can be represented as
\begin{equation}
\text{Selective Feature} = \text{Configuration} \;?\; \text{Data A} \;:\; \text{Data B}.
\end{equation}

\subsection{Modeling the Single-stage Module}
\label{sec:single_module}

\begin{algorithm}[!t]
\caption{Power model of module A in Fig.~\ref{fig:toy_design}}
\label{alg:module_power}
\vspace{1pt}{\color{clrRule}\hrule height 0.7pt}\vspace{1pt}
\begin{lstlisting}[style=pseudoverilog]
// module interface
input  [7:0]  data;              // data signal
input         vld, rdy;          // control signals
input         mask, a_mode;      // config. signals
input  [15:0] coef0, coef1, coef2, bias;  // coefficients
// data-signal post-processing and binning
wire [3:0] data_hd  = HD(data);   (*@\lstcmt // HD: see \eqref{eq:hd}@*)
wire [1:0] data_bin =
      ({2{data_hd >= 6}}                 & 2'd3)
    | ({2{data_hd >= 3 & data_hd <= 5}}  & 2'd2)
    | ({2{data_hd >= 1 & data_hd <= 2}}  & 2'd1);
// multiplicative interaction: control x data
wire [1:0] mult = {2{vld & rdy}} & data_bin;
// selective interaction: config. -> operational modes
wire sel0 = ~a_mode & ~mask;  // MUL,  un-masked
wire sel1 = ~a_mode &  mask;  // MUL,  masked
wire sel2 =  a_mode;          // MAC
// power model -- linear-model style
wire [1:0] prxy0 = {2{sel0}} & mult;
wire [1:0] prxy1 = {2{sel1}} & mult;
wire [1:0] prxy2 = {2{sel2}} & mult;
output [18:0] pwr = coef0*prxy0 + coef1*prxy1
                  + coef2*prxy2 + bias;
// power model -- decision-tree style
wire [15:0] leaf_coef = ({16{sel0}} & coef0)
                      | ({16{sel1}} & coef1)
                      | ({16{sel2}} & coef2);
output [18:0] pwr = leaf_coef * mult + bias;
\end{lstlisting}
\vspace{3pt}{\color{clrRule}\hrule height 0.7pt}
\end{algorithm}

An example power model of the single-stage circuit module~A shown in Fig.~\ref{fig:toy_design} is presented in Algorithm~\ref{alg:module_power}. The algorithm is written in synthesizable pseudo Verilog code. The feature interactions are constructed manually following the previous sub-sections. The coefficients are fit with module~A's simulated power trace.

A single-stage module can be modeled by either a linear model or a decision tree when following our proposed feature construction method. Recall that a plain linear model has the form
\begin{equation}
\hat{y} = \beta_0 + \sum_{j=1}^{p} \beta_j x_j,
\label{eq:linear}
\end{equation}
where $x_j$ is the $j$-th post-processed proxy and $\beta_j$ is its fitted coefficient. A CART (Classification and Regression Trees) decision tree model has the form
\begin{equation}
\hat{y} = \sum_{m=1}^{M} c_m \, \mathbf{1}(\mathbf{x} \in R_m),
\label{eq:cart}
\end{equation}
where $R_1, \dots, R_M$ are proxies interacted and partitioned by tree, $c_m$ is the coefficient of each partition, and $\mathbf{1}(\cdot)$ is the indicator function selecting the activated partition. 

Because the control and configuration signals cleanly separate the operational modes, the two forms describe the same model. The interaction makes the mode indicators \texttt{sel0}--\texttt{sel2} of Algorithm~\ref{alg:module_power} mutually exclusive, so they partition the proxy space exactly as the leaves of \eqref{eq:cart} do. The decision tree selects the coefficient of the active mode with a multiplexer and multiplies it by the power proxies, while the linear model multiplies each mode by a different coefficient and sums the products. Since only one mode is active at a time, both models return the same value and can be used interchangeably. However, in practice the linear model fit with L1 regularization appears to be more robust to noise and outliers than bagged decision trees fit with dropout and pruning. Therefore, the linear model is used as our default for modeling the single-stage modules.

\subsection{Modeling a Complete Design}
\label{sec:complete_design}

\begin{figure}[!t]
\centering
\includegraphics[width=\columnwidth]{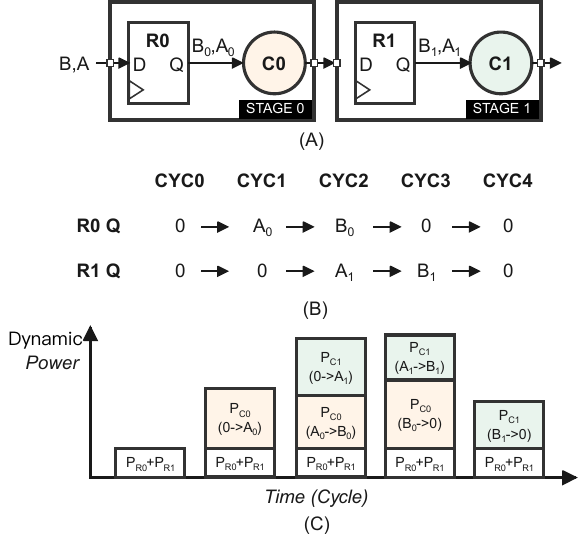}
\caption{Dynamic power fluctuation of a two-stage pipeline. (A) Block Diagram of the pipeline (B) Data toggle at the output of two registers. (C) Per-cycle dynamic power decomposition.}
\label{fig:pipeline}
\end{figure}

RTL designs are usually pipelined, which can be modeled using generalized additive models (GAMs)~\cite{hastie1986gam}. The general form of a GAM is
\begin{equation}
g\bigl(E(y \mid \mathbf{x})\bigr) = \beta_0 + \sum_{i=1}^{p} f_i(x_i),
\label{eq:gam}
\end{equation}
where $x_i$ denotes the \emph{power proxy} of module~$i$, i.e., an RTL signal derived from the three-category method, and $f_i$ denotes the \emph{power signature} of $x_i$, i.e., the dynamic power stimulated when $x_i$ asserts. The intercept $\beta_0$ represents the power that does not vary with the workload, such as leakage power and the power of un-gated registers. The conditional expectation $E(y \mid \mathbf{x})$ together with the link function $g(\cdot)$ relates the toggles in $\mathbf{x}$ to the expected power $y$.

We use an example to explain why pipelined designs can be modeled additively. Fig.~\ref{fig:pipeline} illustrates two combinational circuits $\texttt{C0}$ and $\texttt{C1}$ synchronized through two input registers $\texttt{R0}$ and $\texttt{R1}$. In cycle~0, data $A$ arrives at the input port of $\texttt{R0}$. In cycle~1, $A$ reaches the output port of $\texttt{R0}$ and stimulates $\texttt{C0}$, producing a power pulse $P_{\texttt{C0}}(0 {\to} A_0)$. In cycle~2, data $B$ reaches the output port of $\texttt{R0}$ and stimulates $\texttt{C0}$, producing $P_{\texttt{C0}}(A_0 {\to} B_0)$; meanwhile, $A$ arrives at the output port of $\texttt{R1}$ and stimulates $\texttt{C1}$. Given that $\texttt{R0}$ and $\texttt{R1}$ are always on, the total dynamic power of cycle~2 and cycle~3 is
\begin{align}
P_{\texttt{CYC2}} &= P_{\texttt{C0}}(A_0 {\to} B_0) + P_{\texttt{C1}}(0 {\to} A_1) + P_{\texttt{R0}} + P_{\texttt{R1}},
\label{eq:cyc2}\\
P_{\texttt{CYC3}} &= P_{\texttt{C0}}(B_0 {\to} 0) + P_{\texttt{C1}}(A_1 {\to} B_1) + P_{\texttt{R0}} + P_{\texttt{R1}}.
\label{eq:cyc3}
\end{align}

The above equations demonstrate that the power of a pipelined design is inherently additive. A complete RTL design normally contains a hierarchy of pipelined modules, and is designed to perform multiple functions. GAM can be used to model a complete RTL design in two different ways---the \emph{module-based} and \emph{function-based} style. Under module-based style, each $f_i(x_i)$ represents the power model of a module, which is applicable to most of the RTL designs. In contrast, the function-based style utilizes $f_i(x_i)$ to represent the power model of a function. For example, when modeling a processor, we can use one proxy to fit the power signature of one instruction. The proxy values varying through time are designed to be positively correlated with the processor's dynamic power. Function-based method allows us to build the power model directly from the control plane of an RTL design, which is more hardware efficient for designs containing massive repetitive structures, such as Tensor Cores containing hundreds of identical Multiply-Accumulate (MAC) units.

\subsection{Construct Train and Test Dataset}
\label{sec:dataset}

\begin{table}[!t]
\centering
\caption{Categories of pre-silicon test cases.}
\label{tab:testcases}
\renewcommand{\arraystretch}{1.2}
\setlength{\tabcolsep}{3pt}
\footnotesize
\begin{tabular}{@{}L{0.17\columnwidth}L{0.39\columnwidth}L{0.39\columnwidth}@{}}
\toprule
\textbf{Category} & \textbf{Verification Purpose} & \textbf{Example} \\
\midrule
Direct &
Stimulate a target module or function to check its correctness &
Issue one instruction to check the functional correctness of the processor \\
\addlinespace[3pt]
Random &
Randomize simulation parameters to uncover corner-case bugs &
Randomly issue instructions to check the in-order retirement of an out-of-order processor \\
\addlinespace[3pt]
Benchmark &
Check the architectural model's performance matches with the RTL implementation &
CoreMark~\cite{coremark} for CPUs or a complete neural network inference for GPUs \\
\bottomrule
\end{tabular}
\end{table}

The quality of the training and testing datasets drastically affects the model accuracy. We propose a method to efficiently collect datasets for fitting the model. Recall that a linear model can be used to fit the model built with our proposed modeling method. If perfect feature interaction is performed, all $p$ features ($x_j$) in \eqref{eq:linear} will be orthogonal. If we construct a training dataset that comprises $p$ samples, each having exactly one non-zero feature, then \eqref{eq:linear} can be solved deterministically as
\begin{equation}
\beta=(X^TX)^{-1}X^Ty
\end{equation}
where $y$, $X$, $\beta$ denote the power vector, training dataset, and the fitted coefficients, respectively. In reality, the interacted features will not be perfectly orthogonal, and manually crafting test cases for each proxy in a billion-transistor SoC design is intractable. To bridge this gap, we propose to reuse the \emph{functional test cases} built during the chip development phase for training the power model.

Functional test cases are rigorously engineered to validate the correctness of the RTL design. Industrial standards usually require 90\%+ code coverage and 100\% functional coverage to be achieved before sign-off for RTL verified with functional test cases. As a result, functional test cases usually cover the activation of all modules and the functions of an RTL design. They can be regarded as a superset of the aforementioned ideal training cases. Table~\ref{tab:testcases} summarizes three categories of functional test cases.

We propose to reuse the direct cases and the random cases for training the power model, and to use the benchmark cases for testing the model, instead of mixing all test cases together and dividing them by a ratio. Our method de-correlates the data distributions of the training set and the testing set and helps us identify the generalization problem of many modeling methodologies. If the existing functional test cases are not enough for the model training to converge, supplementary random cases can be engineered. Two criteria should be checked for the training set: (1) the power values of the training dataset must span from leakage power to the maximum power of the RTL design; (2) every proxy must toggle at least once.

\section{Automatic Model Building Procedure}
\label{sec:auto}
While the white-box modeling approach offers superior explainability and robustness, the manual feature interaction step proposed in Section~\ref{sec:postproc} requires prohibitive modeling effort for complex RTL designs. To reduce the modeling effort, we propose algorithms to automatically learn the feature interactions while keeping the meaning of the learned features similar to the ones built with the white-box approach. We also show how to integrate the automatic feature interaction into the full modeling pipeline.

\subsection{Characteristic of the Power Modeling Problem}
Automated, or black-box power modeling approaches~\cite{lin2019ensemble,xie2021apollo,li2025cobit} commonly take two steps. First, a feature selection algorithm such as a linear model with an $L_1$ penalty screens out the candidate power proxies that are most sensitive to power fluctuation from a superset. The superset is typically all nets in the netlist or the input signals of all modules. Second, a regression backbone such as a linear model with an $L_2$ penalty, a decision tree, or a neural network~\cite{isscc2026proactive} is used to fit the power trace collected from power simulation.

Feature interaction, though not frequently referred to in prior art, is the key to transforming the selected raw features so that they fit the characteristics of the regression model. From a statistical learning perspective, different regression models suit features and datasets with different peculiarities. Polynomial models, including linear, spline and other variants, are efficient when a few strong, non-multicollinear features correlate to the target smoothly. Tree models are efficient when a few strong features correlate to the target in a disjointed, conditional and step-like manner, since a decision path is itself a conjunction of thresholds. Neural networks are efficient when the target is driven by many weak, interdependent features through a smooth mapping, and the number of samples is significantly larger than the number of features. 

The characteristics of the raw features and the datasets of on-chip power modeling are as follows. First, the response is not smooth. As shown in Section~\ref{sec:postproc}, power is governed by gated, conditional combinations of signals, so the mapping from toggle rates to power is step-like rather than continuous; these signals are also strongly collinear, since multiple signals may toggle simultaneously when a power event arrives. Second, the strong features are sparse: prior art~\cite{xie2021apollo,li2025cobit} consistently reports that a few hundred features ($<0.1\%$ of the superset) are sufficient to converge the model fitting. Third, the training set is small. The slow netlist power simulation and the requirement to finish the model fitting before sign-off cap the trace at millions to billions of cycles for a complex server-grade processor core, and windowing further reduces the number of samples.

Consequently, we propose to use a decision tree to automatically learn the feature interactions discussed in Section~\ref{sec:postproc}. Our modeling procedure differs from other tree model works~\cite{li2025cobit,lin2019ensemble} in two ways. First, the raw features fed to the feature interaction step have been screened by humans following the rules in Table~\ref{tab:signal_categories}. Second, a procedure is set up to check whether the interacted features follow the white-box principles described in Section~\ref{sec:postproc}. 

\begin{figure}[!t]
\centering
\includegraphics[width=\columnwidth]{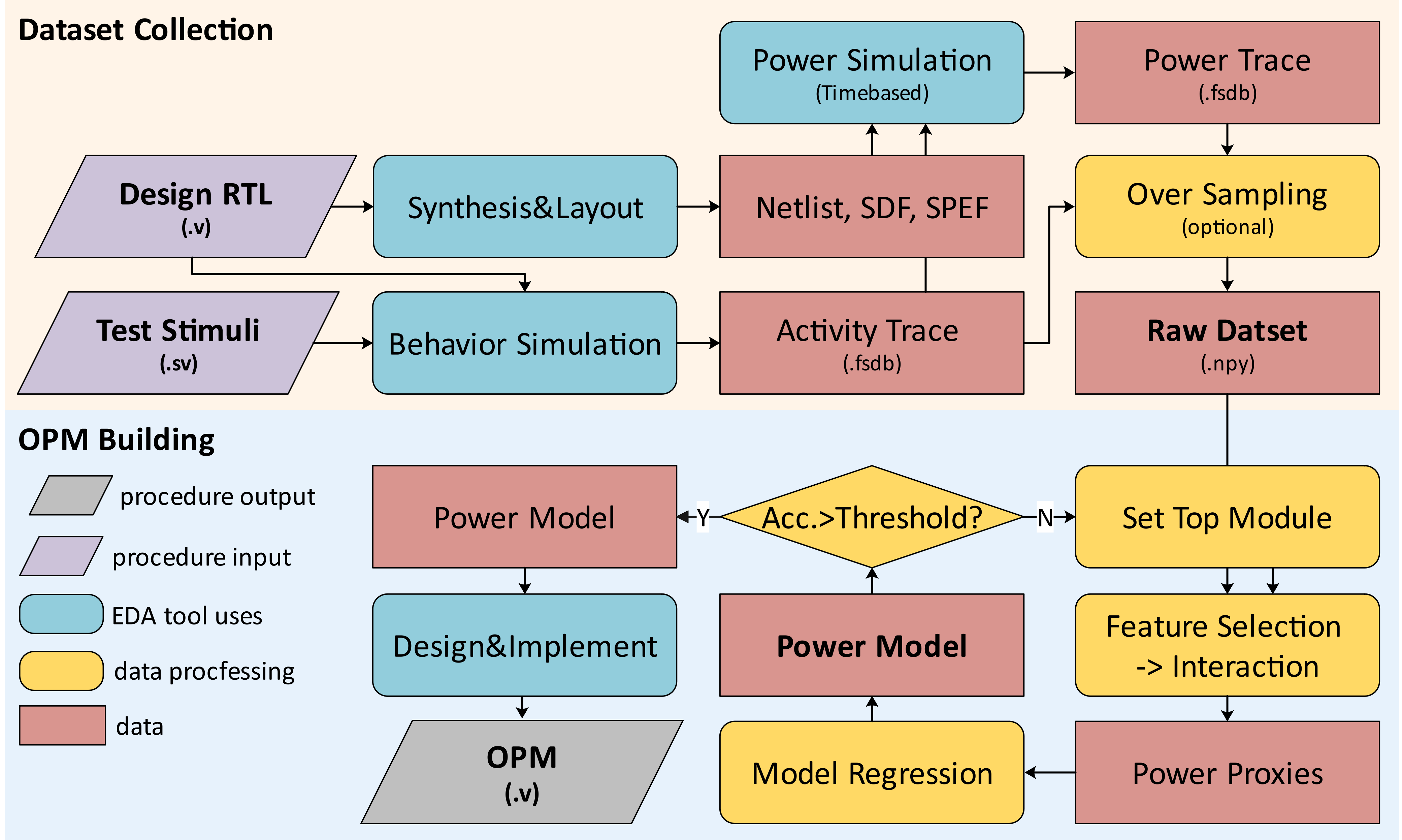}
\caption{The proposed automatic modeling pipeline}
\label{fig:training_flow}
\end{figure}

After the feature interaction step, the modeling procedure remains the same as the white-box modeling procedure proposed in Section~\ref{sec:complete_design}: a linear or tree model is used to fit the single-stage modules, and GAM is used to hierarchically model the complete RTL design. Similar to prior art, feature selection step is added before the feature interaction step to coarsely screen the raw features. Fig.~\ref{fig:training_flow} summarizes the complete OPM modeling pipeline, and we discuss each step in the following subsections.

\subsection{Feature Selection}

Prior works~\cite{xie2021apollo,zhai2022deep,li2025cobit} build OPM models from a superset of RTL or netlist signals, but many of these signals are repetitive (such as signals flopped in a pipeline) or irrelevant to power. The number of signals in the superset can be orders of magnitude larger than the number of signals actually correlated to power. For example, an edge processor core can contain over $10^6$ RTL signals, but using a few hundred of them is sufficient to converge the OPM model fitting~\cite{li2025cobit}.

To quickly discard irrelevant signals, a feature selection step is applied to the signal superset with acceptable computational cost. Prior works usually rely on sparse linear regression constrained with LASSO or MCP (Minimax Concave Penalty)~\cite{xie2021apollo} to perform feature selection. Sparse linear regression penalizes the coefficients of features that are weakly correlated with power. The features are then ranked by the magnitude of their coefficients, and the top $K$ features are selected to enter the next step of the modeling pipeline. A hyperparameter sweep is applied to find the optimal $K$ that balances model accuracy and hardware cost~\cite{li2025cobit}.

We propose a two-step feature selection approach. First, we discard any features exhibiting zero variance across the training set. Then, we apply five selection algorithms in parallel to identify the top $K$ features, and adopt the algorithm that achieves the highest $R^2$ score on the validation set. Hyperparameter optimization (HPO) is employed to tune the hyperparameters of each algorithm, and a timeout mechanism is applied to discard algorithms that fail to converge within a reasonable time. The five algorithms include: the \textit{Variance Threshold} algorithm, which purges signals falling below a specified variance limit; the \textit{Pearson Coefficient} algorithm, which eliminates features lacking sufficient linear correlation with power; the same \textit{MCP} algorithm proposed by~\cite{xie2021apollo}; the \textit{Model-based Selection} algorithm, which trains a statistical model to extract the $K$ most important features; and the \textit{Recursive Feature Elimination (RFE)} algorithm, which iteratively trains a model to prune the weakest signals until exactly $K$ features remain.

\subsection{Feature Interaction and Model Regression}

While feature selection removes signals that are weakly correlated with power, it does not account for interactions among the selected signals. As discussed in Section~\ref{sec:postproc}, selective and multiplicative interactions commonly exist between these retained signals and can be constructed manually. In this section, we introduce an algorithm to automatically generate these feature interactions and fit the model.

COBIT~\cite{li2025cobit} found that decision trees can be used to construct the feature interactions. They recognized that the decision trees construct chaining conditional branch nodes that represent circuit design features, such as the chained clock-enable signals. We follow COBIT's approach and derive deep feature interactions between signals with Gradient Boosted Decision Trees (GBDT). But unlike COBIT, we only use the tree model to propose the possible interacted features instead of using it as the final predictive model. 

Our proposed model fitting flow is described as follows. First, we process the dataset following the approach described in Section~\ref{sec:postproc} to improve the correlation between the signals and power. Second, we feed the processed dataset to a GBDT with a large depth limit to construct the interacted features. We rank the constructed features by feature importance and remove the features whose depth equals $1$. Among the high-importance features, we manually select the interacted features that follow the semantics described in Section~\ref{sec:postproc}. Finally, we feed the selected features together with the processed control signals to the linear model for model fitting.

The GBDT we used can be formulated as follows. Given the processed dataset $\{(\mathbf{x}_i, y_i)\}_{i=1}^{n}$, where $\mathbf{x}_i \in \mathbb{R}^{p}$ is the toggle-rate vector of the $i$-th window and $y_i$ is its power label, the GBDT approximates power as an additive ensemble of $M$ regression trees:
\begin{equation}
F_M(\mathbf{x}) = F_0 + \eta \sum_{m=1}^{M} h_m(\mathbf{x}),
\label{eq:gbdt}
\end{equation}
where $F_0 = \frac{1}{n}\sum_{i=1}^{n} y_i$ is the initial prediction and $\eta \in (0,1]$ is the learning rate. The trees are fitted stagewise. At iteration $m$, the negative gradient of the squared loss, i.e., the residual $r_{im} = y_i - F_{m-1}(\mathbf{x}_i)$, becomes the target of a new tree:
\begin{equation}
h_m = \arg\min_{h} \sum_{i=1}^{n} \left( r_{im} - h(\mathbf{x}_i) \right)^2,
\label{eq:gbdt_fit}
\end{equation}
and the ensemble is updated as $F_m = F_{m-1} + \eta h_m$. Each tree $h_m$ partitions the feature space into $L_m$ disjoint leaf regions $\{R_{lm}\}_{l=1}^{L_m}$ and assigns a constant $\gamma_{lm}$ to each leaf:
\begin{equation}
h_m(\mathbf{x}) = \sum_{l=1}^{L_m} \gamma_{lm} \, \mathbb{I}\left(\mathbf{x} \in R_{lm}\right).
\label{eq:gbdt_tree}
\end{equation}
A leaf region is the conjunction of the split conditions along its root-to-leaf path $\mathcal{P}_{lm}$:
\begin{equation}
\mathbb{I}\left(\mathbf{x} \in R_{lm}\right) = \prod_{(j,\,t,\,s) \in \mathcal{P}_{lm}} \mathbb{I}\left( s \cdot (x_j - t) > 0 \right),
\label{eq:gbdt_rule}
\end{equation}
where $j$, $t$, and $s \in \{-1,+1\}$ denote the split feature, the threshold, and the branch direction, respectively. Eq.~\eqref{eq:gbdt_rule} is exactly the product form of the selective interaction in Section~\ref{sec:postproc}: a path of depth $d$ gates the power contribution by $d$ signals. We therefore extract every path as a candidate interacted feature, whose depth equals $|\mathcal{P}_{lm}|$ and whose importance is the total loss reduction contributed by its splits.

We utilize ElasticNet as the linear model. It blends L1 (Lasso) and L2 (Ridge) regularization to optimize feature selection and numerical stability. The L1 penalty forces aggressive sparsity, crushing weak or physically meaningless proxy coefficients to zero. Simultaneously, the L2 penalty stabilizes multi-collinearity, guaranteeing that highly correlated, physically valid proxy groups are scaled proportionally rather than arbitrarily shattered by the L1 solver. The objective loss function is:
\begin{equation}
Loss = WRSS + \lambda \left( \alpha \sum_{j=1}^{p} |\beta_j| + \frac{1 - \alpha}{2} \sum_{j=1}^{p} \beta_j^2 \right)
\label{eq:loss}
\end{equation}

\subsection{Training Skills}
\label{sec:train_skill}

To optimize peak-power prediction accuracy, the ElasticNet uses Weighted Least Squares (WLS). The Weighted Residual Sum of Squares ($WRSS$) is formulated as:
\begin{equation}
WRSS = \sum_{i=1}^{n} w_i \left( y_i - \hat{y}_i \right)^2
\label{eq:wrss}
\end{equation}
By assigning aggressively scaled non-negative weights $w_i$ to maximum-power vectors, WLS forces the optimizer to prioritize high-di/dt transients, consciously sacrificing low-power idle accuracy to ensure the proactive power management system detects critical anomalies.

During regression, we enforce two physical constraints on the ElasticNet. First, the intercept $\beta_0$ is fixed to the static leakage power. When all input proxies evaluate to zero, the predicted dynamic power must mathematically collapse to baseline leakage. Second, all learned coefficients are strictly bounded to non-negative values. Because dynamic logic toggles physically cannot generate negative power, forcing positive coefficients guarantees that the model maps strictly to physical energy dissipation.

All hyperparameters are tuned automatically by Bayesian optimization with the Tree-structured Parzen Estimator (TPE). Within strict temporal bounds, the TPE iteratively samples configurations, evaluates them against the validation set via $R^2$, and implements early-stopping. For the GBDT stage, we optimize tree depth, estimator count, logarithmic learning rate, and sub-sampling ratios. For the ElasticNet stage, we tune global regularization strength ($\lambda$) and the precise L1/L2 mixing ratio ($\alpha$).

A persistent critical issue in OPM design is severe dataset imbalance. High-power vectors represent a minuscule statistical fraction of standard functional testbenches, as they require massive, synchronized multi-module toggling. Standard least-squares solvers naturally bias toward the overwhelming volume of idle/low-power states. To mathematically rectify this, we deploy ADASYN, an adaptive synthetic oversampling algorithm, to synthetically multiply high-power edge cases, rebalancing the gradient topography prior to regression.

During model synthesis, the top-level architectural block undergoes automated feature selection, GBDT interaction, and ElasticNet regression. The candidate model is aggressively validated against the isolated test-set using $R^2$, RMSE, and sMAPE metrics. If the model fails to satisfy a strict baseline $R^2_{\text{target}}$, the methodology executes a hierarchical descent. The block is shattered into its constituent sub-modules, the entire X-OPM pipeline is executed independently on each child, and the per-module results are summed as in Eq.~\ref{eq:gam}. We heavily utilize a human-in-the-loop intervention layer, permitting hardware designers to manually inject known critical proxies or purge physically meaningless rules. To minimize silicon overhead, any sub-module contributing less than 5\% to the parent's total power budget is deliberately skipped; its power footprint is statically absorbed into the parent's linear intercept.

\begin{figure}[!t]
\centering
\includegraphics[width=\columnwidth]{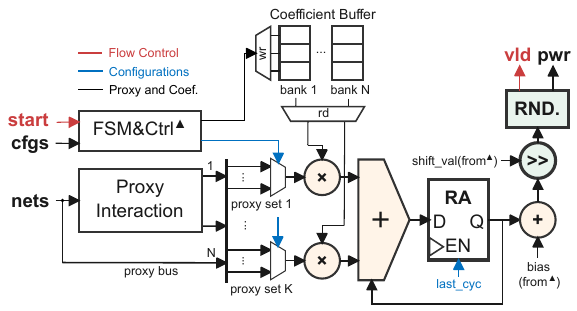}
\caption{Architecture of the OPM predictor and its model evaluation datapath.}
\label{fig:opm_predictor}
\end{figure}

\section{OPM Deployment}
\label{sec:deployment}

\subsection{OPM Hardware Design}

Once an OPM is trained, its proxies and the shape of its model are fixed, and the model is hardened on silicon as the OPM predictor (Fig.~\ref{fig:opm_predictor}). The predictor evaluates
\begin{equation}
\label{eq:hw_model}
\text{pwr} = \mathrm{RND}\Bigl(\bigl(\sum_{i=0}^{N-1} q_i x_i + b\bigr) \gg \mathit{sval}\Bigr)
\end{equation}
where $N$ is the number of retained proxies, $x_i$ is the signal count of the $i$-th proxy accumulated over a \textit{sampling window}, $q_i$ is the fitted non-zero coefficient quantized to 8-bit unsigned integer, and $b$ is the quantized unsigned intercept. The interacted features built by GBDT are hardened inside the proxy interaction module, together with the selected processed nets, form the proxy buses that will be evaluated by the arithmetic path. The arithmetic path of the predictor includes a coefficient buffer that stores the trained coefficients; $K$ multipliers and an adder tree to perform the MAC (multiply-and-accumulate) operation of the linear model in a time-multiplexing fashion; an accumulation register (RA) to gather and accumulate the partial products of the MAC in multiple cycles; an output adder to add the bias; an arithmetic right shifter to statically quantize the result to 8-bit based on a pre-determined shift value ($sval$), as well as to perform averaging; and a rounding module (RND) that performs round to nearest tie to even on the quantized result. The predictor can be fully-pipelined to achieve the same frequency as the processor core. 

The coefficient buffer is designed as a memory implemented with register files instead of fixed combination circuits encoding the trained coefficients. This design is crucial for enabling the OPM to be retrained with post-silicon data after tape-out. 

Time multiplexing is used to reduce the number of multipliers and the scale of the adder tree: OPM usually requires averaging the calculated power in a time window to improve the prediction accuracy. This is because varying latency exists between the toggling of different proxies and the generation of the power. The existence of this time window lets us compute the \eqref{eq:hw_model} in $T$-cycles instead of 1-cycle, where $T$ depends on the training results but can range to around several cycles to tens of several cycles. As a result, we implement $K$ instead of $N$ multipliers. The proxy bus is split into $K$ proxy sets, each owning one multiplexer and one multiplier. In every cycle, the FSM\&Ctrl block steers each multiplexer to one proxy of its set and reads the matching coefficient from the coefficient buffer, so $\lceil N/K \rceil$ cycles are required to calculate one power data. We design $T$ to be a power of $2$ so the averaging is a right shift rather than a division. On the last cycle of the partial sum accumulation, the control signal \texttt{last\_cyc} triggers the accumulation register RA to flop the power data, and feeds it to the downstream modules. 

We prune the learnt features with RFE to reduce more hardware overhead of implementing a trained model. RFE is used to prune the proxies: at each step, we drop the proxy whose coefficient contributes least to the prediction and refit the ElasticNet on the rest. Smaller $N$ yields lower $R^2$, and we choose the smallest $N$ that does not harm the $R^2$.

\subsection{Hardware Overhead Analysis}
\label{sec:ppa}

In this section, we analyze the scalability of the proposed OPM predictor by quantitatively evaluating the relationship between the area scaled with the number of proxies $N$ and the time window $T$. 

We estimate the area in relative units. We consider the area of 1-bit full adder or flip-flop to be $1$ unit; a 1-bit half adder to be $0.75$ unit; a 1-bit 2-to-1 multiplexer and a 2-input XOR gate to be $0.5$ unit; and a 2-input AND and OR gate to be $0.25$ unit. To simplify the calculation, fixed arithmetic circuit structure is used, which includes array multiplier, carry-propagate adder, and the log-shifter. We assume the timing margin is enough so logic inflation and standard cell size-up is negligible. 

Equation \eqref{eq:hw_model} takes $\lceil N/K\rceil$ cycles to finish one prediction with $K$ multipliers, and requires $K\ge\lceil N/T\rceil$. A 1-bit proxy counted over $T$ cycles needs $w_x=\lceil\log_2(T+1)\rceil$ bits, and coefficients are $w_q=8$ bits. An array multiplier then costs $A_{\mathrm{mul}}\approx0.25\,w_qw_x+w_q(w_x-1)$, a tree adder $A_{\mathrm{add}}\approx w_q+w_x$, a coefficient entry $A_{\mathrm{coef}}\approx w_q$, a proxy's counter and hold register $A_{\mathrm{cnt}}\approx2.5\,w_x$, and the select logic of one multiplexed proxy $A_{\mathrm{mux}}=(w_q+w_x)/2$. The total is
\begin{equation}
A \approx K\,(A_{\mathrm{mul}}\!+\!A_{\mathrm{add}}) + N\,(A_{\mathrm{coef}}\!+\!A_{\mathrm{cnt}}) + (N\!-\!K)\,A_{\mathrm{mux}}
\label{eq:area}
\end{equation}
The accumulation register (RA), output accumulator, shifter, RND and FSM\&Ctrl modules are omitted from the area calculation since they do not scale with $N$ and $T$. Proxy interaction module is also omitted since it depends on the GBDT results and normally does not consume a significant amount of area. When taking $K=\lceil N/T\rceil$, the \eqref{eq:area} becomes
\begin{equation}
A(N,T) \approx N(A_{\mathrm{coef}}\!+\!A_{\mathrm{cnt}}\!+\!A_{\mathrm{mux}}) + \Bigl\lceil\frac{N}{T}\Bigr\rceil(A_{\mathrm{mul}}\!+\!A_{\mathrm{add}}\!-\!A_{\mathrm{mux}})
\label{eq:area_nt}
\end{equation}
The first term scales as $O(N\log T)$ since every count and multiplexer widens with $w_x$. The second term scales as $O(N(\log T)/T)$; in sum, we have
\begin{equation}
A(N,T) = O(N\log T).
\label{eq:order}
\end{equation}

The area is thus linear in $N$ and only logarithmic in $T$. At $N=30$, Eq.~\eqref{eq:area_nt} evaluates the area to be $850$, $872$, and $907$ units for $T=4$, $8$, and $16$, a $6.7\%$ spread over a $4\times$ range, so $T$ should be chosen for accuracy rather than area, and increasing $T$ adds at most $T$ cycles of latency. Reducing the number of proxies ($N$) is the most effective way to reduce the hardware overhead. At $T=8$, with $N=372$, the circuit needs $K=47$ multipliers and $10.7$k-unit area; with $N=30$, the circuit only needs $K=4$ multipliers and $872$-unit area, resulting in about $12.3\times$ area reduction.
\section{Experiments}
\label{sec:experiments}

\begin{figure}[!t]
\centering
\includegraphics[width=\columnwidth]{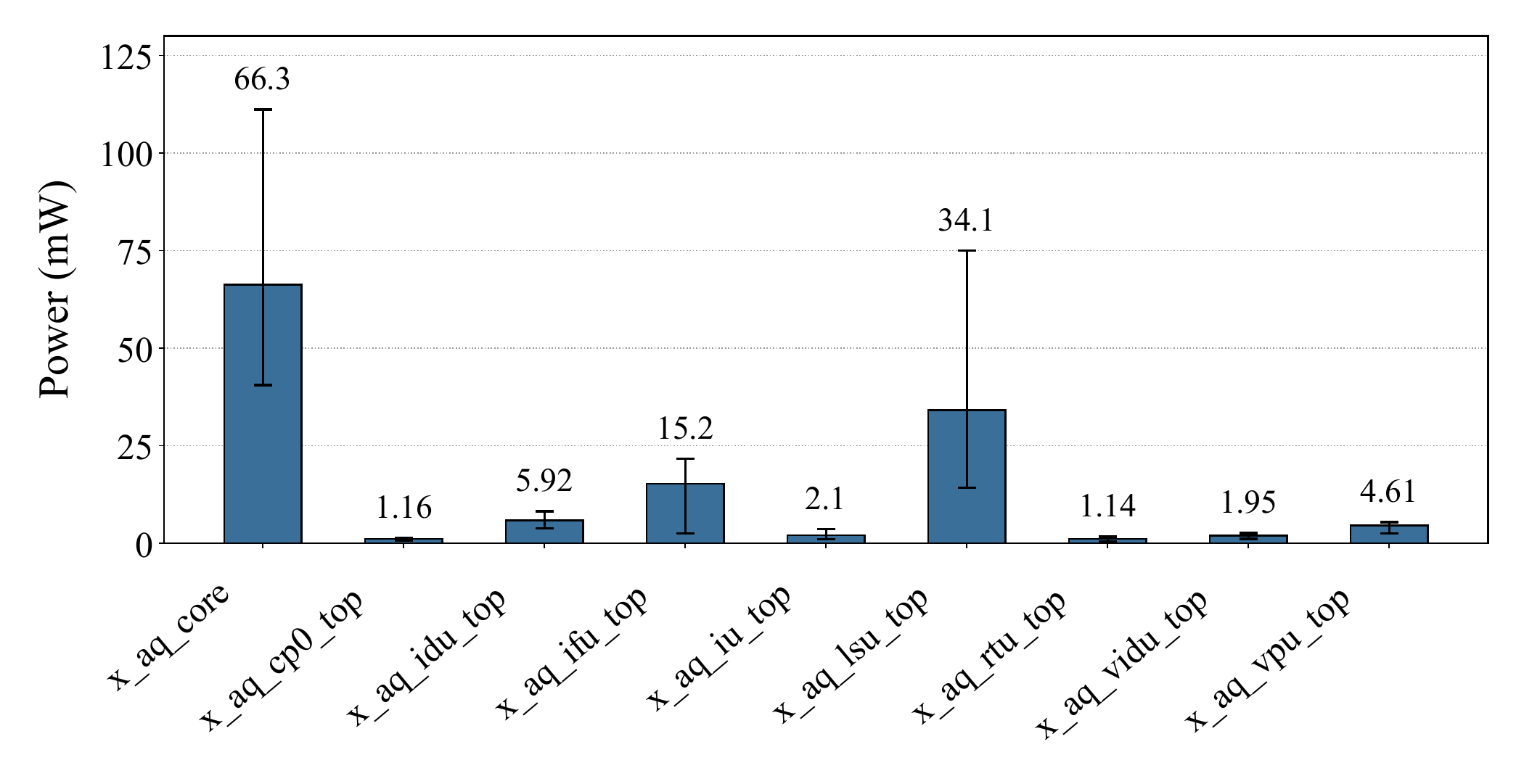}
\caption{Power breakdown of the C906 core running the CoreMark benchmark. Bar heights show average module power; error bars show the power fluctuation range during the execution.}
\label{fig:c906_breakdown}
\end{figure}

\subsection{Experiment Setup}

\begin{figure*}[!t]
\centering
\includegraphics[width=\textwidth]{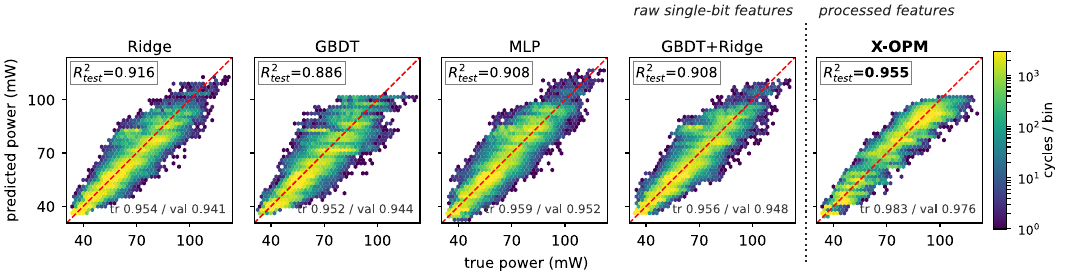}
\caption{The residual plot shows the predicted versus true per-cycle power of \texttt{x\_aq\_core} on the test benchmarks. The four baselines on the left are trained on raw single-bit features; X-OPM is trained on the processed features. Each hexagon bin is colored by the number of cycles it holds; the red dashed line marks perfect prediction. Train and validation $R^2$ are listed in the lower-right corner of each panel.}
\label{fig:comparison_aq_core}
\end{figure*}

We evaluate the effectiveness of X-OPM on the XuanTie C906~\cite{c906datasheet}, an open-sourced RISC-V edge processor developed by Alibaba T-Head. The C906 implements the RV64GCV instruction set, supports RVV 0.7.1 vector extension via a 128-bit vector lane. It supports various data types including INT8, FP16, FP32, etc. It is one of the ideal edge processors for GOPS-level neural network inference at the edge. The C906 is a mass-produced, silicon-proven core supported by production-grade software stacks, including the customized GCC compiler, the HHB (Heterogeneous Honey Badger) AI compiler and the ShiHuLan high-performance operator library.

We synthesized and laid out the C906 core using TSMC 28\,nm HPC+ technology node. Industrial standard tools, including the Synopsys Design Compiler 2025 and the Cadence Innovus 21.1 are used to implement the C906. We perform post-layout simulation with the time resolution of 1 cycle to extract the power data of the benchmarks. Synopsys PrimePower 2025 was used to perform the time-based power simulation. Data extraction was performed on HPC cluster with hundreds of processing cores and terabyte level of RAMs.

The layout of C906 core consumes approximately 0.87\,mm$^2$ area with a 42\% standard cell density. Timing is closed at 1\,GHz under 0.9\,V supply voltage TT corner. Our implementation shares the supply voltage rail of the logic and the SRAM, and utilizes a mixture of ULVT, LVT and SVT cells for implementation. 

\subsection{Modeling Result of C906}

Fig.~\ref{fig:c906_breakdown} shows the simulated power distribution of the C906 core executing CoreMark, which is an industrial-standard MCU benchmark including matrix multiplication, dynamic list traversal, and deep state-machine, etc. Each bar in the figure represents a hardware module under the C906 core top, which is marked as \texttt{x\_aq\_core}. Other modules, suffixed with \texttt{\_top}, are sub-modules under \texttt{x\_aq\_core}.

We modeled the OPM of C906 as the GAM of all sub-modules under the C906 core top, except that \texttt{x\_\allowbreak aq\_\allowbreak lsu\_\allowbreak top} and \texttt{x\_\allowbreak aq\_\allowbreak vpu\_\allowbreak top} are modeled as the GAM of their own sub-modules to improve their accuracy. Sub-modules that generate less than $5\%$ of their parent module's power are skipped to simplify the power modeling procedure.

Fig.~\ref{fig:comparison_aq_core} compares X-OPM with four baselines on the core-level power of \texttt{x\_aq\_core}. The baselines are Ridge regression, GBDT, an MLP, and a GBDT+Ridge hybrid, and all four take the raw single-bit signals as features. X-OPM uses the same regression backbone as GBDT+Ridge but takes the processed features of Section~\ref{sec:postproc}. X-OPM reaches a test $R^2$ of $0.955$, while the best baseline, Ridge, reaches $0.916$. The more expressive baselines do not close this gap: GBDT reaches $0.886$ and MLP $0.908$, although both fit the training set as well as Ridge does. The drop from validation to test $R^2$ is also smallest for X-OPM ($0.021$, against $0.025$--$0.058$ for the baselines), so the processed features generalize to unseen workloads rather than memorizing the training benchmarks. The GBDT+Ridge result isolates the source of the gain: with raw features it scores $0.908$, and replacing only the features raises it to $0.955$. The remaining error of X-OPM concentrates at the high-power end, where its predictions saturate near $100$\,mW. These peak cycles are rare in the training set, a limitation we return to in Section~\ref{sec:discussion}.

\subsection{Hardware Overhead Evaluation}
\label{sec:hw_overhead}

\begin{figure}[!t]
\centering
\includegraphics[width=\columnwidth]{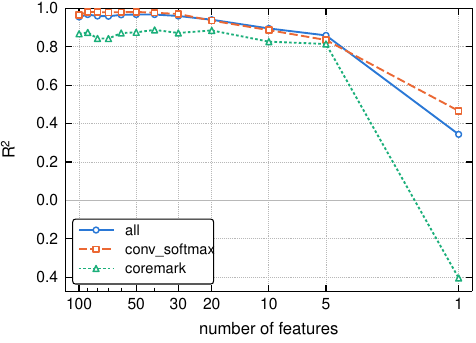}
\caption{Test $R^2$ of the C906 OPM versus the number of retained proxies under RFE pruning.}
\label{fig:xopm_prune}
\end{figure}

Fig.~\ref{fig:xopm_prune} shows the RFE feature pruning results after the model is trained. Our trained model originally contains $372$ proxies; the resulting $R^2$ varying against $N$ for the first $100$ proxies is shown in the figure. We explicitly show the influence of the pruning on different test cases. Accuracy is flat from $100$ down to about $30$ proxies, where the pooled test $R^2$ stays near $0.89$. Below $20$ proxies it starts to decline: the pooled $R^2$ falls to $0.83$ at $N=10$ and $0.79$ at $N=5$. With a single proxy, the pooled $R^2$ drops to $0.28$, and CoreMark becomes negative $-0.47$, meaning the model predicts worse than the workload mean. We deploy the pruned model with $30$ proxies. 

After layout, the OPM predictor with the pruned model occupies $5307\,\mu$m$^2$, which is $0.61\%$ of the $0.87\,$mm$^2$ C906 core.

\subsection{Discussion about the Modeling Results}
\label{sec:discussion}

The results above leave two open questions: why X-OPM saturates at the high-power end, and why more expressive models do not remove that error. Both follow from how the training data is generated and what least-squares fitting does with it.

\textit{Peak power is rare in the training data.} A peak-power cycle, and the $di/dt$ spike that accompanies it, requires several events to coincide: a dense instruction mix, wide datapath toggling, and deep combinational paths that are all un-gated in the same cycle. Ordinary workloads seldom produce this alignment. Stress vectors that do produce it are also the most expensive to simulate with PTPX, because simulation time grows with switching activity. The training set is therefore dominated by average-power cycles.

\textit{Least-squares fitting learns the average case.} The coefficient $\beta_i$ of each proxy is fit to minimize the error over the whole dataset, so it converges to the mean power drawn when that proxy is active. Lasso and Ridge shrink the coefficients but do not change this behavior. The two proxy types are affected differently. A clock-gating signal is active in almost every workload, so its coefficient is well estimated and describes average-case power. A data-gating signal is deasserted mainly in high-power cycles, so its coefficient depends on the few peak samples in the dataset and is estimated poorly. The saturation near $100$\,mW in Fig.~\ref{fig:comparison_aq_core} is the visible result.

\textit{Model capacity does not substitute for features.} OPM modeling combines four properties: many candidate signals, a non-smooth mapping from signals to power, near-zero label noise because PTPX is deterministic, and few independent workloads. In this regime, a high-capacity model can fit the training benchmarks exactly without learning the circuit structure that transfers to new workloads. The baseline comparison in Fig.~\ref{fig:comparison_aq_core} supports this: GBDT and the MLP fit the training set as well as Ridge but score lower on the test set. Features derived from the design's gating structure, fit by a regularized linear model, generalize better than raw signals fit by a larger model.

\textit{Accuracy scales with modeling effort and hardware cost.} A model built only on top-level clock gates captures coarse average power. Descending the RTL hierarchy and modeling individual sub-modules improves accuracy, as done for \texttt{x\_aq\_lsu\_top} and \texttt{x\_aq\_vpu\_top}. Fine-grained configuration and data signals, however, add proxies and therefore OPM logic. They should be reserved for the sub-modules whose power is both large and variable, which is also where the peak-power error concentrates.

\section{Conclusion}
\label{sec:conclusion}

In this paper, we introduced X-OPM, an explainable, automated modeling methodology for building accurate digital on-chip power meters with low hardware overhead. By utilizing feature engineering developed according to the low-power design rules of the synchronous digital VLSI circuits, X-OPM mitigates the robustness problem met by the prior black-box modeling methodologies. X-OPM proposed a white-box feature interaction methodology, then proposed an automatic modeling framework to implement it. X-OPM proposed to utilize tree models to extract the explainable feature interactions, and utilize a regularized linear model to perform the model regression. The results of the tree model are reviewed by humans to guarantee it obeys the rules set by the X-OPM's white-box methodology. Evaluated on an open-source but silicon-proven XuanTie C906 RISC-V processor with TSMC 28\,nm process and commercial EDA tools, X-OPM achieved a test $R^2$ of $0.955$ on unseen benchmarks, outperforming Ridge, GBDT, MLP, and GBDT+Ridge baselines trained on raw signals, which reached $0.886$--$0.916$. The deployed predictor occupies less than $0.1\%$ core area after layout. In sum, X-OPM proposes a robust, white-box oriented OPM modeling methodology and framework for building OPMs that remain trustworthy after tape-out.

\bibliographystyle{IEEEtran}
\bibliography{references}

\begin{thebibliography}{10}
\providecommand{\url}[1]{#1}
\csname url@samestyle\endcsname
\providecommand{\newblock}{\relax}
\providecommand{\bibinfo}[2]{#2}
\providecommand{\BIBentrySTDinterwordspacing}{\spaceskip=0pt\relax}
\providecommand{\BIBentryALTinterwordstretchfactor}{4}
\providecommand{\BIBentryALTinterwordspacing}{\spaceskip=\fontdimen2\font plus
\BIBentryALTinterwordstretchfactor\fontdimen3\font minus
  \fontdimen4\font\relax}
\providecommand{\BIBforeignlanguage}[2]{{%
\expandafter\ifx\csname l@#1\endcsname\relax
\typeout{** WARNING: IEEEtran.bst: No hyphenation pattern has been}%
\typeout{** loaded for the language `#1'. Using the pattern for}%
\typeout{** the default language instead.}%
\else
\language=\csname l@#1\endcsname
\fi
#2}}
\providecommand{\BIBdecl}{\relax}
\BIBdecl

\bibitem{grochowski2002didt}
E.~Grochowski, D.~Ayers, and V.~Tiwari, ``Microarchitectural simulation and
  control of di/dt-induced power supply voltage variation,'' in \emph{Proc.
  Int. Symp. on High-Performance Computer Architecture (HPCA)}, Feb. 2002.

\bibitem{powell2002damping}
M.~D. Powell and T.~N. Vijaykumar, ``Pipeline damping: A microarchitectural
  technique to reduce inductive noise in supply voltage,'' in \emph{Proc. Int.
  Symp. on Computer Architecture (ISCA)}, May 2002.

\bibitem{joseph2003voltage}
R.~Joseph, D.~Brooks, and M.~Martonosi, ``Control techniques to eliminate
  voltage emergencies in high performance processors,'' in \emph{Proc. Int.
  Symp. on High-Performance Computer Architecture (HPCA)}, Feb. 2003.

\bibitem{leng2014gpuvolt}
J.~Leng, Y.~Zu, M.~Rhu, M.~Gupta, and V.~J. Reddi, ``{GPUVolt}: Modeling and
  characterizing voltage noise in {GPU} architectures,'' in \emph{Proc. Int.
  Symp. on Low Power Electronics and Design (ISLPED)}, Jul. 2014.

\bibitem{leng2015gpunoise}
J.~Leng, Y.~Zu, and V.~J. Reddi, ``{GPU} voltage noise: Characterization and
  hierarchical smoothing of spatial and temporal voltage noise interference,''
  in \emph{Proc. Int. Symp. on High-Performance Computer Architecture (HPCA)},
  Feb. 2015.

\bibitem{thomas2016emergpu}
R.~Thomas, K.~Barrera, J.~Leng \emph{et~al.}, ``{EmerGPU}: Understanding and
  mitigating resonance-induced voltage noise in {GPU} architectures,'' in
  \emph{Proc. Int. Symp. on Performance Analysis of Systems and Software
  (ISPASS)}, May 2016.

\bibitem{paul2013cooperative}
I.~Paul, S.~Manne, M.~Arora, W.~L. Bircher, and S.~Yalamanchili, ``Cooperative
  boosting: Needy versus greedy power management,'' in \emph{Proc. Int. Symp.
  on Computer Architecture (ISCA)}, Jun. 2013.

\bibitem{ibmz15proactive}
C.~Berry \emph{et~al.}, ``Proactive power management in {IBM} z15,'' \emph{IBM
  Journal of Research and Development}, Aug. 2020.

\bibitem{qualcomm2021hexagon}
S.~Bang \emph{et~al.}, ``A proactive system for voltage-droop mitigation in a
  7-nm {Hexagon} processor,'' \emph{IEEE Journal of Solid-State Circuits}, Feb.
  2021.

\bibitem{lefurgy2017power9}
C.~Lefurgy \emph{et~al.}, ``Adaptive clocking in the {POWER9} processor for
  voltage droop protection,'' in \emph{Proc. IEEE Int. Solid-State Circuits
  Conf. (ISSCC)}, Feb. 2017.

\bibitem{ibmz14power}
C.~Berry \emph{et~al.}, ``{IBM} z14: Processor characterization and power
  management for high-reliability mainframe systems,'' \emph{IEEE Journal of
  Solid-State Circuits}, Jan. 2019.

\bibitem{proactive65nm2024}
X.~Chen, J.~Feng, A.~Shoukry, X.~Zhang, R.~Magod, N.~V. Desai, and J.~Gu, ``A
  65-nm proactive power management technique with real-time machine learning
  engine for droop prediction and mitigation on microprocessors,'' \emph{IEEE
  Journal of Solid-State Circuits}, vol.~60, no.~6, pp. 2170--2181, Jun. 2025.

\bibitem{ondie2025runtime}
Y.-L. Chen \emph{et~al.}, ``Run-time power management system by on-die power
  sensor with silicon machine learning-based calibration in a 3-nm mobile
  {SoC},'' in \emph{Proc. IEEE Int. Solid-State Circuits Conf. (ISSCC)}, Feb.
  2025.

\bibitem{isscc2026proactive}
J.~Kim \emph{et~al.}, ``Proactive power management-based supply regulation with
  online learning for variation-tolerant workload-aware processors,'' in
  \emph{Proc. IEEE Int. Solid-State Circuits Conf. (ISSCC)}, Feb. 2026.

\bibitem{xie2021apollo}
Z.~Xie, X.~Xu, M.~Walker, J.~Knebel, K.~Palaniswamy, N.~Hebert, J.~Hu, H.~Yang,
  Y.~Chen, and S.~Das, ``{APOLLO}: An automated power modeling framework for
  runtime power introspection in high-volume commercial microprocessors,'' in
  \emph{Proc. IEEE/ACM Int. Symp. on Microarchitecture (MICRO)}, Oct. 2021.

\bibitem{zhai2022deep}
J.~Zhai, Y.~Cai, and B.~Yu, ``{DEEP}: Developing extremely efficient runtime
  on-chip power meters,'' in \emph{Proc. IEEE/ACM Int. Conf. on Computer-Aided
  Design (ICCAD)}, Oct. 2022.

\bibitem{li2025cobit}
Q.~Li, W.~Lyu, Y.~Liu, J.~Tao, and J.~Han, ``Coordinating binary trait:
  Accurate and lightweight runtime on-chip power meter design,'' \emph{IEEE
  Trans. on Very Large Scale Integration (VLSI) Systems}, vol.~33, no.~12, pp.
  3369--3382, Dec. 2025.

\bibitem{c906datasheet}
{T-Head Semiconductor}, ``Xuantie c906 processor datasheet,''
  \url{https://occ-oss-prod.oss-cn-hangzhou.aliyuncs.com/resource//1718100372780/XuanTie%20C906%20Processor%20Datasheet_20240607.pdf},
  2024.

\bibitem{li2009mcpat}
S.~Li, J.~H. Ahn, R.~D. Strong, J.~B. Brockman, D.~M. Tullsen, and N.~P.
  Jouppi, ``{McPAT}: An integrated power, area, and timing modeling framework
  for multicore and manycore architectures,'' in \emph{Proc. IEEE/ACM Int.
  Symp. on Microarchitecture (MICRO)}, Dec. 2009.

\bibitem{zhou2019primal}
Y.~Zhou, H.~Ren, Y.~Zhang, B.~Keller, B.~Khailany, and Z.~Zhang, ``{PRIMAL}:
  Power inference using machine learning,'' in \emph{Proc. ACM/IEEE Design
  Automation Conf. (DAC)}, Jun. 2019.

\bibitem{zhang2020grannite}
Y.~Zhang, H.~Ren, and B.~Khailany, ``{GRANNITE}: Graph neural network inference
  for transferable power estimation,'' in \emph{Proc. ACM/IEEE Design
  Automation Conf. (DAC)}, Jul. 2020.

\bibitem{lin2019ensemble}
Z.~Lin, S.~Sinha, and W.~Zhang, ``An ensemble learning approach for in-situ
  monitoring of {FPGA} dynamic power,'' \emph{IEEE Trans. on Computer-Aided
  Design of Integrated Circuits and Systems}, vol.~38, no.~9, pp. 1661--1674,
  Sep. 2019.

\bibitem{prophetplus2024}
J.~Fang \emph{et~al.}, ``Data-pattern-based predictive on-chip power meter in
  {DNN} accelerator,'' \emph{IEEE Trans. on Computer-Aided Design of Integrated
  Circuits and Systems}, Nov. 2024.

\bibitem{hastie1986gam}
T.~Hastie and R.~Tibshirani, ``Generalized additive models,'' \emph{Statistical
  Science}, vol.~1, no.~3, pp. 297--310, 1986.

\bibitem{coremark}
{EEMBC}, ``{CoreMark}: An {EEMBC} benchmark,''
  \url{https://www.eembc.org/coremark/}, 2009.

\end{thebibliography}

\end{document}